\documentclass{IEEEtran}
\usepackage[utf8]{inputenc}
\usepackage{amssymb}
\usepackage[ruled,vlined]{algorithm2e}

\SetCommentSty{mycommfont}
\usepackage{float}
\usepackage{balance}

\usepackage{algpseudocode}
\usepackage{booktabs} %%Tables
\usepackage{graphicx} %%Tables
\usepackage[table,xcdraw]{xcolor} %%Tables
\usepackage{multirow, hhline} %%Tables
\usepackage{caption}
\usepackage{subcaption}
\usepackage{dblfloatfix}
\usepackage{amsmath}
\usepackage{amsfonts}
\usepackage{enumitem}
\usepackage{xcolor}

\newcommand{\Hy}{{\cal H}}
\newcommand{\V}{{\cal V}}
\newcommand{\N}{{\cal N}}

\newcommand{\Pins}[1]{{\tt pins}[#1]\xspace}
\newcommand{\Nets}[1]{{\tt nets}[#1]\xspace}
\newcommand{\Part}{{{\tt part}}\xspace}
\newcommand{\Parts}{{{\tt parts}}\xspace}

\usepackage{lipsum}

\newcommand{\mmref}{{\sc REF}\xspace}
\newcommand{\mmrefrlx}{{\sc REF\_RLX}\xspace}
\newcommand{\mmrefrlxsv}{{\sc REF\_RLX\_SV}\xspace}

\title{Streaming Hypergraph Partitioning Algorithms on Limited Memory Environments}
\author{\IEEEauthorblockN{Fatih Taşyaran\IEEEauthorrefmark{1}, Berkay Demireller\IEEEauthorrefmark{1}, Kamer Kaya\IEEEauthorrefmark{1} and Bora Uçar\IEEEauthorrefmark{2}}\\
\IEEEauthorblockA{\IEEEauthorrefmark{1}Computer Science and Engineering, Sabancı University, İstanbul, Turkey\\
{\tt \{fatihtasyaran, bdemireller, kaya\}@sabanciuniv.edu}}\\
\IEEEauthorblockA{\IEEEauthorrefmark{2}CNRS and LIP, (UMR5668 Univ. de Lyon, ENS Lyon, Inria, UCBL 1), France\\
{\tt bora.ucar@ens-lyon.fr}}
}

\begin{document}
\date{}
\maketitle
\begin{abstract}
Many well-known, real-world problems involve dynamic data which describe the relationship among the entities. Hypergraphs are powerful combinatorial structures that are frequently used to model such data. For many of today's data-centric applications, this data is streaming; new items arrive continuously, and the data grows with time. With paradigms such as Internet of Things and Edge Computing, such applications become more natural and more practical. {In this work, we assume a streaming model where the data is modeled as a hypergraph, which is generated at the edge. This data then partitioned and sent to remote nodes via an algorithm running on a memory-restricted device such as a single board computer.} {Such a partitioning is usually performed by taking a {\em connectivity} metric into account to minimize the communication cost of later analyses that will be performed in a distributed fashion.} Although there are many {\em offline} tools that can partition static hypergraphs excellently, algorithms for the streaming settings are rare. We analyze a well-known algorithm from the literature and significantly improve its running time by altering its inner data structure. For instance, on a medium-scale hypergraph, the new algorithm reduces the runtime from $17800$ seconds to $10$ seconds. We then propose sketch- and hash-based algorithms, as well as ones that can leverage extra memory to store a small portion of the data to enable the refinement of partitioning when possible. {We experimentally analyze the performance of these algorithms and report their run times, connectivity metric scores, and memory uses on a high-end server and four different single-board computer architectures.}
\end{abstract}

\section{Introduction}
Real-world data can be complex and there can be natural and irregular relations within, which makes most of the models such as column- or row-oriented tabular representation fail in capturing the essence of knowledge contained. Hypergraphs, which are generalizations of graphs, are highly flexible and appropriate for such data. Simply being a set of sets,  a hypergraph is widely used in various areas to model the data at hand. For instance, they are employed for DNA sequencing~\cite{Venkatraman_2018}, VLSI design~\cite{vlsi}, citation recommentation~\cite{Onur12}, and finding semantic similarities between documents ~\cite{menezes2019semantic} or descriptor similarities between images~\cite{Skiker2014}. They are also used in machine learning since their inherent structure makes them an easy way of representing labels and attributes~\cite{hglabel}. 

Distributed graph and hypergraph stores became popular in the last decade to store massive-scale data that we have in today's applications. A good partitioning of the data among the nodes in the distributed framework is necessary to reduce the communication, i.e., data transfer, overhead of the further analyses. With the increasing popularity of data-centric paradigms such as Internet of Things and Edge Computing, the data that is fed to these stores started to have a streaming fashion. In this work, we assume that the data is generated/processed at the edge of a network and partitioned on a memory-restricted device such as single-board computers (SBC). There exist many algorithms to partition streaming graphs~\cite{akin, fennel, ldg}, and two recent benchmarks to evaluate the performance of such algorithms~\cite{bench1, bench2}. Although hypergraphs tend to have a better modeling capability, and we have fine-tuned, optimized, fast offline hypergraph partitioning tools, e.g.,~\cite{andre2017memetic,patoh,umpa,vlsi}, the streaming setting is not analyzed thoroughly in the context of hypergraph partitioning. 

Hypergraphs are considered as a generalization of graphs; in a graph, the edges represent pairwise connections. On the other hand, in a hypergraph, connectivity is modeled with nets which can represent multi-wise connections. That is, a single net connects more than two vertices. In a streaming setting, this difference makes the hypergraph partitioning problem is much harder than its graph counterpart. For graph partitioning, when a vertex appears with its edges, the endpoint vertex IDs are implicit; hence, just the part information of the vertices is sufficient to judiciously decide on the part of the vertex at hand. However, in a hypergraph, a vertex appears with its nets and the neighbor vertices are not implicit. Hence, one needs to keep track of the connectivity among the nets and the parts to judiciously decide the part of the current vertex. 

In this work, we assume a streaming setting where the vertices of a hypergraph appear in some order along with their nets, i.e., their connections to other vertices. The contribution of the study is three-fold: 
\begin{itemize}[leftmargin=*]
\item We take one of the existing and popular algorithms from the literature~\cite{alistarh2015min-max} and make it significantly faster by altering its inner data structure used to store the part-to-net connectivity. 
\item We propose techniques to refine the existing partitioning at hand with the help of some extra memory to store some portion of the hypergraph. 
\item We propose and experiment with various algorithms and benchmark their run times, memory usages, and partitioning quality on a high-end server and multiple SBCs.
\end{itemize}

The rest of the paper is organized as follows: Section~\ref{sec:not} presents the notation and background on streaming hypergraph partitioning. The proposed algorithms are presented in Section~\ref{sec:alg}. The related work is summarized in Section~\ref{sec:rel}. Section~\ref{sec:exp} presents the experiments  and Section~\ref{sec:con} concludes the paper.

\section{Notation and Background}\label{sec:not}

A hypergraph $\Hy = (\V,\N)$ is composed of a vertex set $\V$
and a net (hyperedge) set $\N$ connecting the vertices in $\V$.  A net $n
\in \N$ is also a vertex set and the vertices in $n$ are the {\sl pins\/} of $n$.
The {\em size} of $n \in \N$ is the number of pins in it, and the {\em degree} of $v \in V$ is the number of nets it
is connected to.  The notation $\Pins{n}$ and $\Nets{v}$ represent the pin set of
a net $n$, and the set of nets containing a vertex $v$, respectively.
In this work, we assume that the vertices and the nets are homogeneous, i.e., they all have equal weights and costs. However, in practice, vertices can be associated with weights, and nets can be associated with costs.

A {\sl $K$-way partition\/} of $\Hy$, which is denoted as $\Pi\!=\!\{\V_1,\V_2, \ldots
,\V_K\}$, is a vertex partition where
\begin{itemize}[leftmargin=*]
\item parts are pairwise disjoint, i.e., $\V_k \cap \V_\ell =
  \emptyset$ for all $1 \leq k < \ell \leq K$,
\item each part $\V_k$ is a nonempty subset of $\V$, i.e., $\V_k
  \subseteq \V$ and $\V_k \neq \emptyset$ for $1\leq k \leq K$,
\item the union of $K$ parts is equal to $\V$, i.e., $\bigcup_{k=1}^K \V_k
  \!=\!\V$. 
\end{itemize}

A {\sl $K$-way partition\/} of $\Hy$, which is denoted as $\Pi\!=\!\{\V_1,\V_2, \ldots
,\V_K\}$, is a vertex partition where
\begin{itemize}[leftmargin=*]
\item parts are pairwise disjoint, i.e., $\V_k \cap \V_\ell =
  \emptyset$ for all $1 \leq k < \ell \leq K$,
\item each part $\V_k$ is a nonempty subset of $\V$, i.e., $\V_k
  \subseteq \V$ and $\V_k \neq \emptyset$ for $1\leq k \leq K$,
\item the union of $K$ parts is equal to $\V$, i.e., $\bigcup_{k=1}^K \V_k
  \!=\!\V$. 
\end{itemize}

In the streaming setting, the vertices in $\V$ appear one after another. 
The elements of the stream are pairs ($v, \Nets{v}$). 
For each element, $v$ is be partitioned, i.e., the part vector entry $\Part[v]$ is set by the partitioning algorithm. In the strict streaming setting, each element ($v, \Nets{v}$) is forgotten after $\Part[v]$ is decided. 
Besides, none of the partitioning decisions can be revoked. 
In a more flexible streaming setting, a buffer with a capacity $B$ is reserved to store some of the net sets. These vertices can then be re-processed and re-partitioned. In this setting, the cost of storing  ($v, \Nets{v}$) in the buffer is $|\Nets{v}|$.

At any time point of the partitioning, the partition must be kept {\em balanced} 
by limiting the difference between the number of vertices inside
the most loaded and the least loaded part. 
Let $s$ be the slack denoting this limit. We say that the partition is balanced 
if and only if $$abs(|\V_i| - |\V_j|) \leq s \mbox{ for all } 1 \leq i < j \leq K$$ where $abs(x)$ is the absolute value of $x$.
For a $K$-way partition $\Pi$, a net that has at least one pin
(vertex) in a part is said to {\sl connect\/} that part.  The number
of parts connected by a net $n$ is called its {\sl connectivity\/} and
denoted as $\lambda_n$.  A net $n$ is said to be {\sl uncut\/} ({\sl
internal}) if it connects exactly one part (i.e., $\lambda_n = 1$),
and {\sl cut\/} ({\sl external}), otherwise (i.e., $\lambda_n >
1$). Given a partition $\Pi$, if a vertex is in the pin set of at
least one cut net, it is called a {\it boundary vertex}.

In the text, we use
$\Parts[n]$ to denote the set of parts net $n$ is connected to. Let
$\Lambda(n,p) = |\Pins{n} \cap \V_p|$ be the number of pins of net $n$
in part $p$. Hence, $\Lambda(n,p) > 0$ if and only if $p \in $
$\Parts[n]$. There are various metrics to measure the quality of a partitioning in terms of the connectivity of the nets~\cite {leng:90}. 
The one which is widely used in the literature and shown
to accurately model the total communication volume of many data-processing kernels is called the {\sl connectivity-1} metric. This cutsize metric is defined as:
 
\begin{align}
\chi(\Pi) &=  \sum_{n \in {\N}}(\lambda_n - 1)\; \label{eq:con}
\end{align}

\noindent In this metric, each cut net $n$ contributes
\mbox{$(\lambda_n - 1)$} to the cut size. The hypergraph
partitioning problem can be defined as the task of finding a balanced
partition $\Pi$ with $K$ parts such that $\chi(\Pi)$ is
minimized. This problem is NP-hard even in the offline setting~\cite{leng:90}, where all the vertices can be considered to appear at once, and the balance requirement is only tested at the end of partitioning.

\section{Partitioning Streaming Hypergraphs }\label{sec:alg}

The simplest partitioning algorithm one can employ in the streaming setting is random partitioning, {\sc Random}, which assigns each appearing vertex to a random part while keeping the partitioning always balanced as shown in Algorithm~\ref{alg:ran}. In the algorithm, $p$ is the candidate part, $p_{min}$ is the part ID having the least number of vertices, and $rand(1, K)$ chooses a random integer in between $[1, K]$. When the difference between the number of vertices is equal to $s$, $v$ cannot be assigned to $p$ since this decision makes the partitioning unbalanced.\looseness=-1

\renewcommand{\baselinestretch}{0.9}
\begin{algorithm}[htbp]
\caption{{\sc Random}}\label{alg:ran}
\textbf{Input:} $\Hy$, $K$, $s$\\
\textbf{Output:} $\Part[.]$\\
\For{all \textnormal{(}$v, \Nets{v}$\textnormal{)} in streaming order} {
    $p \gets rand(1, K)$\\
    \While{$|\V_p| - |\V_{p_{min}}| = s$}  {
        $p \gets rand(1, K)$\\
    }
    $\Part[v] \gets p$\\
    \If{$p = p_{min}$} {
        Update $p_{min}$\\
    }
}
{\bf return} {$\Part$}\\
\end{algorithm}
 \renewcommand{\baselinestretch}{1}

\subsection{Min-Max Partitioning}
{\sc MinMax} is a well-known approach proposed for streaming hypergraph partitioning~\cite{alistarh2015min-max}. The approach, whose pseudocode is given in Algorithm~\ref{alg:minmax}, keeps track of net connectivity, i.e., which net is connected to which part, by leveraging a net set ${\tt p2n}[i]$ for each $1 \leq i \leq K$. Each streaming vertex $v$ is assigned to the part $p$ with the largest intersection set ${\tt p2n}[p] \cap \Nets{v}$ that does not disturb the partitioning being balanced. The idea is simple; every intersecting net will not incur an additional cost after setting $\Part[v]$ to $p$. Hence, the maximum intersection cardinality will yield the best possible greedy partitioning decision. The downside of this implementation is that there is no way of knowing if any of $v$'s nets are connected to a part $i$ without checking ${\tt p2n}[i]$. This approach requires many unnecessary checks since all parts need to be considered even if most of them are not connected to the vertex. Furthermore, when $K$ is large, which is the case for many real-life applications, the problem is exacerbated. Overall, even when the cost of the intersection computation is $\mathcal{O}(1)$ per net, the algorithm takes $\mathcal{O}(K \times |\Hy|)$ where $|\Hy|$ is the number of pins in the hypergraph. which is not acceptable since $K$ can be in the order of thousands, and $|\Hy|$ can be easily in between $10^{9}$ and $10^{10}$ for streaming, massive-scale hypergraphs.\looseness=-1

\renewcommand{\baselinestretch}{0.90}
\begin{algorithm}[htbp]
\caption{{\sc MinMax}}\label{alg:minmax}
\textbf{Input:} $\Hy$, $K$, $s$\\
\textbf{Output:} $\Part[.]$\\
\For{$i$ from $1$ to $K$} {
    ${\tt p2n}[i] \gets \emptyset$
}
\For{all \textnormal{(}$v, \Nets{v}$\textnormal{)} in streaming order} {
    $saved \gets -1$\\
    \For{$i$ from $1$ to $K$} {
        \If{$|\V_i| - |\V_{p_{min}}| < s$}  {
            \If{$|{\tt p2n}[i] \cap \Nets{v}| > saved$} {
                $saved \gets |{\tt p2n}[i] \cap \Nets{v}|$\\
                $p \gets i$\\
            }
        }
    }
    $\Part[v] \gets p$\\
    ${\tt p2n}[p] \gets {\tt p2n}[p] \cup \Nets{v}$\\
    \If{$p = p_{min}$} {
        Update $p_{min}$\\
    }
}
{\bf return} {$\Part$}
\end{algorithm}

\begin{algorithm}[htbp]
\caption{{\sc MinMax-n2p}}\label{alg:minmax-n2p}
\textbf{Input:} $\Hy$, $K$, $s$\\
\textbf{Output:} $\Part[.]$\\
$save[.] \gets$ an array of size $K$\\
$mark[.] \gets$ an array of size $K$ with all $-1$s\\
$pids[.] \gets$ an array of size $K$ \\
$indx[.] \gets$ an array of size $K$ \\

\For{all \textnormal{(}$v, \Nets{v}$\textnormal{)} in streaming order} {
    $active \gets 0$\\
    \For{$n \in \Nets{v}$} {
        \If{$n$ appears for the first time} {
            ${\tt n2p}[n] \gets \emptyset$
        }
        \For{$i \in {\tt n2p}[n]$} {
            \If{$mark[i] \neq v$} {
                $mark[i] \gets v$\\
                $active \gets active + 1$\\
                $pids[active] \gets i$\\  
                $save[active] \gets 1$\\
                $indx[i] \gets active$\\
            } \Else{
                $save[indx[i]] \gets save[indx[i]] + 1$\\
            }
        }
    }

    $saved \gets -1$\\
    \For{$j$ from $1$ to $active$} {
        $i \gets pids[j]$\\
        \If{$|\V_i| - |\V_{p_{min}}| < s$}  {
            \If{$save[j] > saved$} {
                $saved \gets save[j]$\\
                $p \gets i$\\
            }
        }
    }
    $\Part[v] \gets p$\\
    
    \lFor{$n \in \Nets{v}$} {
        ${\tt n2p}[n] \gets {\tt n2p}[n] \cup \{p\}$
    }
    \lIf{$p = p_{min}$} {
        Update $p_{min}$
    }
}
{\bf return} {$\Part$}
\end{algorithm}
\renewcommand{\baselinestretch}{1}

\subsection{Using Net-to-Part Information and Finding Active Parts}

Algorithm~\ref{alg:minmax} needs to go over all the parts one by one which creates a performance problem especially when the number of parts is large. This happens since the connectivity information among the nets and the parts is stored from the parts' perspective. However, if this information had been organized from the perspective of the nets it would be possible to identify the {\em active} parts that are connected to at least one net of the current vertex $v$ at hand. Algorithm~\ref{alg:minmax-n2p} describes the pseudocode of this approach. For an efficient computation, it uses four auxiliary arrays, $save$, $mark$, $pids$, and $indx$. Each of these arrays is of size $K$. However, they are only initialized once, and no expensive reset operation with complexity $\Theta(K)$ is performed after a vertex is processed. Thanks to these arrays, when a part is first identified to be connected to one of the nets in $\Nets{v}$, it is marked to save a single net and placed into the active part array. Once it is placed, the next access to the same part (but for a different net) will only increase the savings of this part. Both these operations can be performed in constant time and no loop over all the parts is required.\looseness=-1 

{\sc MinMax} as proposed in~\cite{alistarh2015min-max}, and as described in Algorithm~\ref{alg:minmax}, has been used in the literature to benchmark novel algorithms for scalable hypergraph partitioning, e.g.,~\cite{mayer2018hype,billionscale}. For instance, it is reported that a hypergraph with $0.43$M vertices and $180$M pins is partitioned into $K = 128$ parts in around $1000$ seconds~\cite{mayer2018hype}. On the other hand, the variant in Algorithm~\ref{alg:minmax-n2p} can partition a hypergraph with $1.1$M vertices and $228$M pins into $K = 2048$ parts in around 200 seconds. 

Considering $K$ is at most in the order of tens of thousands, the extra memory usage due to the four additional arrays is not exhaustive. On the other hand, both {\sc MinMax} and {\sc MinMax-n2p} use approximately the same amount of memory to store the connectivity information. That is the total number of entries in ${\tt n2p}$[.] and ${\tt p2n}$[.] arrays are the same, and exactly $n$ more than the ({\em connectivity} - 1) metric given in Eq.~\ref{eq:con}. This being said, {\sc MinMax-n2p} uses slightly more memory since unlike $K$, $|\N|$ is not known beforehand in the streaming setting, and a dynamic data structure with more overhead is required to organize the connectivity as in ${\tt n2p}$ instead of ${\tt p2n}$.\looseness=-1

\subsection{{\sc MinMax} Variants Using Less Memory}

Based on the hypergraph and the number of parts, the memory usage of the two previous approaches can be overwhelming. For streaming data, there is no upper limit to this cost, and for memory-restricted devices such as SBCs, this can be highly problematic. Here we investigate {\sc MinMax} alternatives that can use a fixed amount of memory.\looseness=-1 

\subsubsection{{\sc MinMax-L}$\ell$}
As explained above, both in Algorithm~\ref{alg:minmax} and Algorithm~\ref{alg:minmax-n2p}, the total memory spent for ${\tt n2p}$ grows as long as the data is streaming. To avoid this, while working similar to Algorithm~\ref{alg:minmax-n2p}, {\sc MinMax-L}$\ell$ restricts the maximum {\em length} of each ${\tt n2p}[.]$ to $\ell$. When a new part $p$ is being added to a ${\tt n2p}[n]$ for a net $n \in \N$, if $|{\tt n2p}[n]| = \ell$, a random part id from ${\tt n2p}[n]$ is chosen and replaced with $p$. Although the connectivity information is only kept in a lossy fashion, we expect it to guide the partitioning decisions for sufficiently large $\ell$ values.\looseness=-1 

\renewcommand{\baselinestretch}{0.9}
\begin{algorithm}[htbp]
\caption{{\sc MinMax-L}$\ell$}\label{alg:minmax-l}
\textbf{Input:} $\Hy$, $K$, $s$, $\ell$\\
\textbf{Output:} $\Part[.]$\\

$\cdots$ same as Algorithm~\ref{alg:minmax-n2p}\\

\For{all \textnormal{(}$v, \Nets{v}$\textnormal{)} in streaming order} {
$\cdots$ same as Algorithm~\ref{alg:minmax-n2p}\\

    \For{$n \in \Nets{v}$} {
        \If{$|{\tt n2p}[n]| < \ell$} {
            ${\tt n2p}[n] \gets {\tt n2p}[n] \cup \{p\}$\\
        } \ElseIf{$p \notin {\tt n2p}$} {
            $idx \gets rand(1, \ell)$\\
            ${\tt n2p}[idx] \gets p$\\
        }
    }
    \If{$p = p_{min}$} {
        Update $p_{min}$\\
    }
}
{\bf return} {$\Part$}
\end{algorithm}
\renewcommand{\baselinestretch}{1}

\subsubsection{{\sc MinMax-BF}}
Bloom filters~(BF) are memory-efficient and probabilistic data structures used to answer whether an element is a member of a set~\cite{10.1145/362686.362692}. Compared to the traditional data structures such as arrays, sets, or hash tables used for the same task, a BF occupies much less space while allowing false positives. If the item is a member of the set the sketch always answers positively. Otherwise, if the item is not in the set, it answers negatively with high probability. However, it also can answer positively in this case. 

A Bloom Filter, which employs an $m$-bit sequence, uses $k$ hash functions to find the indices of bits and set them to $1$ to mark the existence of a new. To answer a query for an item $x$, it simply checks the corresponding $k$ hash functions on $x$, and answer positively if each of the corresponding bits is $1$. An important parameter for a BF is its false positive probability which quantifies the quality of its answers. Assuming the hash functions are independent of each other, when $n$ items are inserted into a BF, $kn$ bits are altered. Hence, the probability of a bit stays zero is $\left(1-1/m\right)^{kn} \approx e^{-kn/m}$, and the false-positive probability can be computed as $\left(1-e^{-kn/m}\right)^{k}$.

The BF variant of {\sc MinMax} runs along the same lines with Algorithm~\ref{alg:minmax}. However, instead of ${\tt p2n}$, it leverages a Bloom Filter {\tt BF} to store connectivity tuples $(n, p)$ which means that the net $n$ has a pin at part $p$. For a given ($v, \Nets{v}$), it goes over all the parts, and for each net in $\Nets{v}$, it queries the corresponding tuple within the {\tt BF}. Then it chooses the part with the most number of positive answers. The pseudocode of this approach is given in Algorithm~\ref{alg:minmax-bf}. As in {\sc MinMax-L}$\ell$, using a BF limits and fixes the amount of memory that will be used to store the connectivity among the nets and the parts.\looseness=-1

\renewcommand{\baselinestretch}{0.9}
\begin{algorithm}[htbp]
\caption{{\sc MinMax-BF}}\label{alg:minmax-bf}
\textbf{Input:} $\Hy$, $K$, $s$, $m$\\
\textbf{Output:} $\Part[.]$\\
${\tt BF} \gets \emptyset$ (creates an $m$-bit, all zero sequence)\\
\For{all \textnormal{(}$v, \Nets{v}$\textnormal{)} in streaming order} {
    $saved \gets -1$\\
    \For{$i$ from $1$ to $K$} {
        \If{$|\V_i| - |\V_{p_{min}}| < s$}  {
            $saved_i \gets 0$\\
            \For{$n \in \Nets{v}$} {
                \If{${\tt BF}.{\tt query}((n, i))$} {
                    $saved_i \gets saved_i + 1$\\
                }
            }
    
            \If{$saved_v > saved$} {
                $saved \gets saved_i$\\
                $p \gets i$\\
            }
        }
    }
    $\Part[v] \gets p$\\
    \For{$n \in \Nets{v}$} {
        ${\tt BF}.{\tt insert}((n, p))$\\
    }
    \If{$p = p_{min}$} {
        Update $p_{min}$\\
    }
}
{\bf return} {$\Part$}
\end{algorithm}
\renewcommand{\baselinestretch}{1}

\subsubsection{{\sc MinMax-MH}}
An approach fundamentally different than {\sc MinMax-L}$\ell$ and {\sc MinMax-BF} is completely forgetting the connectivity information and try to cluster similar vertices with similar $\Nets{.}$ sets into the same parts. A natural tool for this task is hashing; we employ a MinHash-based approach~\cite{minhash}, {\sc MinMax-MH}, to find the part ID for a given vertex. For implementation, we used $k$ hash functions in the form of $h_i(x) = a_ix + b_i \bmod q$ where $1 \leq i \leq k$, $q$ is a prime number, $a_i$ and $b_i$ are random integers chosen from $[0, q)$ per hash function. Given ($v, \Nets{v}$), this approach first computes $(\alpha_1, \alpha_2, \ldots, \alpha_k)$ where $\alpha_i = min_{n \in \Nets{v}}\{h_i(n)\}.$ Then the part ID for $v$ is computed as $$p = \left(\prod_{i =1}^k \alpha_i\right) \bmod K.$$ If $v$ cannot be assigned to $p$ due to the balanced partitioning restriction the approach tries the next part in line, i.e., $p + 1 \bmod K$ until a suitable part is found. 

It is intuitive to think that vertices with similar net sets will end up with closer hash values which will lead to them being put in the same part. The fact that there is no need for additional memory to keep net-part connectivity unlike the previous algorithms makes this approach suitable for low memory environments.\\

\subsection{Buffering and Refining}
Although revoking the partitioning decisions is impossible for the strict streaming setting, in which the net sets are forgotten after the corresponding vertex is put to a part, with an additional buffer to keep the $\Nets{.}$ sets, one can revisit the buffered vertices and put them to a different part if it is good for the ({\em connectivity} - 1) metric. A high-level description of this approach is given in Algorithm~\ref{alg:minmax-n2p-ref}.

\renewcommand{\baselinestretch}{0.9}
\begin{algorithm}[htbp]
\caption{{\sc MinMax-n2p-Ref}}\label{alg:minmax-n2p-ref}
\textbf{Input:} $\Hy$, $K$, $s$, $B$, $passes$ \\
\textbf{Output:} $\Part[.]$\\
$\cdots$ same as Algorithm~\ref{alg:minmax-n2p}\\
${\tt BUF} \gets \emptyset$ (an empty buffer that can store $B$ pins)\\
\For{all \textnormal{(}$v, \Nets{v}$\textnormal{)} in streaming order} {
   $\cdots$ same as Algorithm~\ref{alg:minmax-n2p}\\
   \If{${\tt isBufferable}$\textnormal{(}$v$\textnormal{)}} {
       ${\tt BUF}.{\tt insert}$($v$)\\
       \If{${\tt BUF}.{\tt isFull}$\textnormal{()}} {
            \For{$i$ from $1$ to $passes$} {
                \For{$u \in {\tt BUF}$} {
                    Compute $leaveGain$ for $u$\\
                    \If{${\tt isMoveable}$\textnormal{(}$u$\textnormal{)}} {
                        Remove $u$ from $\Part[u]$\\
                        Find the best part $p$ for $u$\\
                        $\Part[u] \gets p$\\
                        \For{$n \in \Nets{u}$} {
                            ${\tt n2p}[n] \gets {\tt n2p}[n] \cup \{p\}$
                        }
                        \If{$p = p_{min}$} {
                            Update $p_{min}$\\
                        }                   
                    }
                }
            }
            ${\tt BUF} \gets \emptyset$
        }
   }
}
{\bf return} {$\Part$}
\end{algorithm}
\renewcommand{\baselinestretch}{1}

Algorithm~\ref{alg:minmax-n2p-ref} works along the same lines with {\sc MinMax-n2p}. In addition to finding the part id for a vertex $v$, after $v$ is processed it can be chosen to be inserted to the buffer ${\tt BUF}$. Once the buffer is full, the algorithm goes over all the vertices in the buffer $passes$ times. For each vertex $u$, first the $leaveGain$ of $u$ is computed which is the change in the ({\em connectivity} - 1) metric when $u$ is removed from $\Part[u]$. Then if $u$ is decided to be movable, it is removed from $\Part[u]$. A new part $p$ is then found and $u$ is moved to $p$. We employed three strategies, namely \mmref, \mmrefrlx and \mmrefrlxsv, for {\sc MinMax-n2p-Ref}, which differ on how they behave for the functions ${\tt isBufferable}$($.$) and ${\tt isMoveable}$($.$):\looseness=-1

\begin{itemize}[leftmargin=*]
    \item The first strategy \mmref buffers every vertex but finds a new best part $p$ if and only if the corresponding $leaveGain > 0$. That is it only modifies $\Part$ when it is probable to reduce the {{\em connectivity} - 1} metric. Hence, it is a restricted strategy while exploring the search space.
    \item The second strategy \mmrefrlx buffers every vertex and finds a new best part $p$ for all the vertices in ${\tt BUF}$. Hence, compared to the previous one it is a {\em relaxed} strategy. 
    \item The third strategy \mmrefrlxsv only buffers the vertices with small net sets and finds a new best part $p$ for all the vertices in ${\tt BUF}$. It aims to reduce the overhead of refining while keeping its gains still on the table. 
\end{itemize}

To refine, i.e., to compute $leaveGain$, one also needs to keep track the number of pins of each net residing at each part. This almost doubles the memory requirement of the refinement heuristics compared to {\sc MinMax}, since for every a connectivity entry stored in ${\tt n2p}$ , an additional positive integer is required. That is the original entry shows that ``{\em net $n$ is connected to part $p$}", and the additional entry required for refinement shows that ``{\em net $n$ has $k$ pins in part $p$}". The refinement-based algorithms require this information since when $k = 1$, they can detect a gain on the connectivity. 

\section{Related Work}\label{sec:rel}
There exist excellent offline hypergraph partitioners in the literature such as PaToH~\cite{patoh} and HMetis~\cite{vlsi}. Recently, more tools are developed focusing on different aspects and using different approaches: for instance,  Deveci~et~al.~\cite{umpa} focuses on handling multiple communication metrics at once, Mayer~et~al.~\cite{mayer2018hype} focuses on the speed, and Schlag~et~al.~\cite{shhmss2016alenex} focuses more on the quality by using a more advanced refinement mechanism.  

Faraj~et~al.~\cite{faraj2021buffered} recently proposed a streaming graph partitioning algorithm which yields high quality partitioning on streaming graphs utilizing buffering model. In addition, Jafari~et~al.~\cite{JAFARI2021140} proposed a fast parallel algorithm which processes the given graph in batches. The streaming setting has not been analyzed thoroughly {on hypergraphs} except the work by Alisarth et al.~\cite{alistarh2015min-max} which proposes the original {\sc MinMax} algorithm at once. {In this work, we make this algorithm significantly faster by altering its inner data structure used to store the part-to-net connectivity. Furthermore, we propose techniques to refine the existing partitioning at hand with the help of some extra memory to store some portion of the hypergraph.}

\begin{table}[htbp]
\centering
\resizebox{.47\textwidth}{!}{%
\begin{tabular}{@{}
>{\columncolor[HTML]{FFFFFF}}l rrrrr@{}}
\multicolumn{1}{l}{\cellcolor[HTML]{C0C0C0}Matrix} & \multicolumn{1}{l}{\cellcolor[HTML]{C0C0C0}Size} & \multicolumn{1}{l}{\cellcolor[HTML]{C0C0C0}Pins} & \multicolumn{1}{l}{\cellcolor[HTML]{C0C0C0}Max. Deg} & \multicolumn{1}{l}{\cellcolor[HTML]{C0C0C0}Avg. Deg} & \multicolumn{1}{l}{\cellcolor[HTML]{C0C0C0}Deg. Var.} \\ \midrule
{\tt coPapersDBLP} & 0.5M & 30.5M & 3.2K & 56.41 & 66.23 \\ 
{\tt hollywood2009} & 1.1M & 227.8M & 11.4K & 99.91 & 271.69 \\ 
{\tt soc-LiveJournal1} & 4.8M & 68.9M & 13.9K & 14.23 & 42.30 \\ 
{\tt com-Orkut} & 3.0M & 234.3M & 33.3K & 76.28 & 153.92 \\ 
{\tt uk-2005} & 39.4M & 936.3M & 1.7M & 23.72 & 1654.56 \\ 
%{\tt sk-2005} & 50.6M & 1.9B & 8.5M & 38.49 & 2027.92 \\ 
{\tt webbase2001} & 118.1M & 1.0B & 816.1K & 8.63 & 141.79 \\ 
\end{tabular}%
}
\caption{\small{Hypergraphs used for the experiments.}}
\label{tab:graph_shape}
\end{table}

\section{Experimental Results}\label{sec:exp}

We tested the proposed algorithms on hypergraphs created from six matrices downloaded from the SuiteSparse Matrix Collection\footnote{{https://sparse.tamu.edu/}}. The properties of these graphs are given in Table~\ref{tab:graph_shape}. For each $n \times n$ matrix, we create a column-net hypergraph $\Hy = (\V, \N)$ where the vertices~(nets) in $\V$~(in $\N$) correspond to the rows~(columns) of the matrix. Moreover, we tested the algorithms on a cutting-edge server and multiple SBCs. {Since the variants studies in this work have different time-memory and quality tradeoff characteristics, we used a set of single-board computers to observe their performance on SBCs having a different number of cores and different amounts of memory.} The specifications of these architectures are as follows:
\begin{itemize}[leftmargin=*]
    \item \textbf{Server:} Intel Xeon Gold 6140, 2.30GHz, 256GB RAM, gcc 5.4.0, Ubuntu 16.04.6.
    \item \textbf{LattePanda:} Intel Atom x5-Z8350, 1.44GHz, 4096MB DDR3L @ 1066 MHz RAM, gcc 5.4.0, Ubuntu 16.04.2
    \item \textbf{Pi:} Broadcom BCM2837, 1.2GHz, 1024MB LPDDR2 @ 400 MHz RAM, gcc 6.3.0, Raspbian 9
    \item \textbf{Odroid:} ARM Cortex-A15, 2GHz and Cortex-A7 @1.3GHz, 2048MB LPDDR3 @ 933 MHz, 2048MB LPDDR3 @ 933 MHz, gcc 7.5.0, Ubuntu 18.04.1
    %\item \textbf{Jetson:} ARM Cortex-A57, 1.43GHz, 4096MB LPDDR4 @ 1600 MHz, gcc 7.5.0, Ubuntu 18.04.4
\end{itemize}

We implemented algorithms in \texttt{C++} and compiled on each device separately with the above-mentioned \texttt{gcc} version. For each matrix, we created three random streams with different vertex orderings. Although it is an offline partitioner and is not suitable for the streaming setting, we employed PaToH v3.3~\cite{patoh} to evaluate the quality and performance of the streaming algorithms. To force a partitioning being balanced, we used a dynamic slack variable computed by using a constant {\em allowed imbalance ratio} $\beta = 0.1$, for which values in the range $5\%$-$10\%$ are common in the partitioning literature. That is while partitioning the $i$th vertex, the allowed slack is set to $s = {\tt max}(1, \beta \times (i/K))$ which corresponds to the $\beta$ of the average part weight at any time during partitioning.

\begin{figure*}[htbp]
    \centering
    \includegraphics[width=0.8\textwidth]{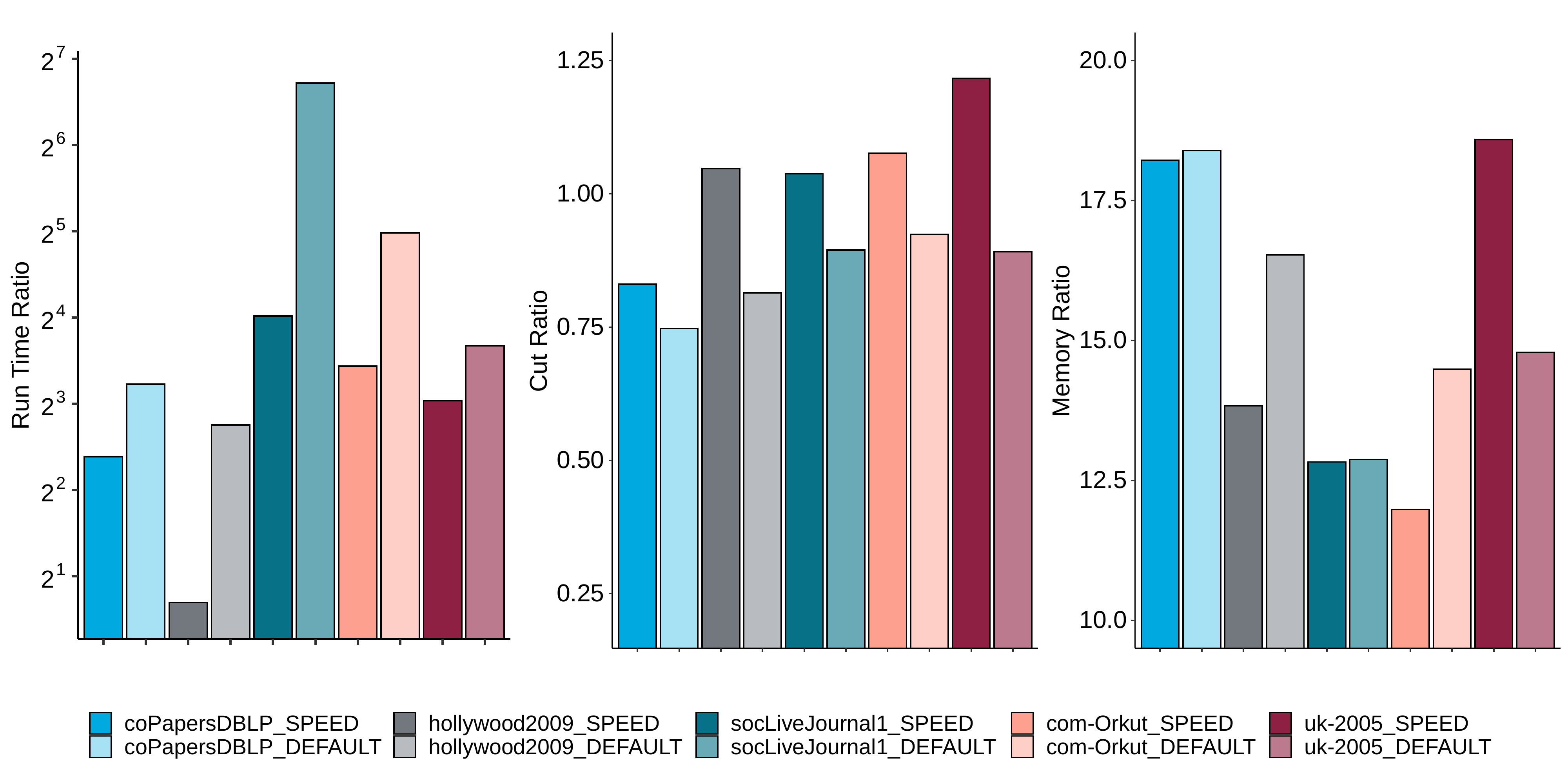}
    \caption{\small{Run times, cuts and memory usages of PaToH normalized with respect to those of {\sc MinMax-n2p}. The experiments are executed on the {\bf Server}.}}
    \label{fig:mmew_comp}
\end{figure*}

\begin{table}[htbp]
\centering
\resizebox{.45\textwidth}{!}{%
\begin{tabular}{@{}llrrr@{}}
\rowcolor[HTML]{9B9B9B} 
\multicolumn{5}{l}{\cellcolor[HTML]{9B9B9B}Parts} \\ \midrule
\rowcolor[HTML]{C0C0C0} 
\multicolumn{1}{|l}{\cellcolor[HTML]{C0C0C0}256} & Algorithm & \multicolumn{1}{l}{\cellcolor[HTML]{C0C0C0}Run Time(sec)} & \multicolumn{1}{l}{\cellcolor[HTML]{C0C0C0}Cut($\times 10^6$)} & \multicolumn{1}{l|}{\cellcolor[HTML]{C0C0C0}Memory(MB)} \\ \midrule
 & {{\sc MinMax}} & {13952.12} & {1.731} & {18.60} \\ \cmidrule(l){2-5} 
 & {{\sc MinMax-n2p}} & {7.40} & {1.731} & {36.78} \\ \cmidrule(l){2-5} 
 & {{\sc MinMax-L}3} & {2.20} & {2.543} & {15.00} \\ \cmidrule(l){2-5}
 & {{\sc MinMax-L}5} & {2.60} & {1.913} & {25.41} \\ \cmidrule(l){2-5}
 & {\sc MinMax-BF}(4) & 496.90 & 3.280 & 2.64 \\ \cmidrule(l){2-5} 
 & {\sc MinMax-BF}(16) & 1479.19 & 3.273 & 2.54 \\ \cmidrule(l){2-5} 
 & {\sc MinMax-MH}(4) & 1.49 & 11.214 & 0.18 \\ \cmidrule(l){2-5} 
 & {\sc MinMax-MH}(16) & 5.37 & 13.464 & 0.19 \\ \cmidrule(l){2-5} 
 & {\sc Random} & 0.04 & 23.817 & 0.00 \\ \midrule
\rowcolor[HTML]{C0C0C0} 
\multicolumn{1}{|l}{\cellcolor[HTML]{C0C0C0}2048} & Algorithm & \multicolumn{1}{l}{\cellcolor[HTML]{C0C0C0}Run Time(sec)} & \multicolumn{1}{l}{\cellcolor[HTML]{C0C0C0}Cut($\times10^6$)} & \multicolumn{1}{l|}{\cellcolor[HTML]{C0C0C0}Memory(MB)} \\ \midrule
 & {{\sc MinMax}} & {17823.45} & {3.096} & {28.65} \\ \cmidrule(l){2-5} 
 & {{\sc MinMax-n2p}} & {10.64} & {3.096} & {44.10} \\ \cmidrule(l){2-5} 
 & {{\sc MinMax-L}3} & {2.61} & {5.217} & {15.00} \\ \cmidrule(l){2-5} 
 & {{\sc MinMax-L}5}  & {3.25} & {4.168} & {25.10} \\ \cmidrule(l){2-5}  
 & {\sc MinMax-BF}(4) & 3664.48 & 4.508 & 2.59 \\ \cmidrule(l){2-5} 
 & {\sc MinMax-BF}(16) & 11882.30 & 4.655 & 2.57 \\ \cmidrule(l){2-5} 
 & {\sc MinMax-MH}(4) & 1.51 & 15.134 & 0.12 \\ \cmidrule(l){2-5} 
 & {\sc MinMax-MH}(16) & 5.38 & 18.101 & 0.15 \\ \cmidrule(l){2-5} 
 & {\sc Random} & 0.07 & 28.992 & 0.00 \\ \cmidrule(l){2-5} 
\end{tabular}%
}
\caption{\small{Comparison of the proposed streaming hypergraph partitioning algorithms and state-of-the-art  on the {\bf Server} and {\tt coPapersDBLP}.}}
\label{tab:versusbad}
\end{table}

Table~\ref{tab:versusbad} reports the run times, cut, and memory usage of the proposed {\sc MinMax} variants on the graph {\tt CoPapersDBLP}. The first two rows show the effectiveness of the modification on {\sc MinMax}. The modified version {\sc MinMax-n2p} is $\approx1700x$ faster on this graph while using $16$--$18$MB more memory. This is due to the reduced number of unnecessary operations on unrelated parts while computing the part savings. The {\sc MinMax-L}3 and {\sc MinMax-L}5 variants are producing partitionings comparable to  {\sc MinMax} in terms of quality~(i.e., with respect to {\em connectivity}-1) while using less memory. 

For {\sc MinMax-BF} and {\sc MinMax-MH}, we use $4$ and $16$ hash functions, and for the former, we use 20M bits. Although being fast, the partitioning quality of the MinHash variant is half of the {\sc Random}. On the contrary, the Bloom Filter variant seems to work well with a comparable partitioning quality. However, it is slow since when ${\tt BF}$ is used instead of ${\tt n2P}$, one needs to go over all the parts to compute their savings. Still, we believe that it is very promising in terms of memory/quality trade-off and enables a scenario in which a device with a small memory is partitioning a vertex stream on a network edge.

Figure~\ref{fig:mmew_comp} compares {\sc MinMax-n2p}'s performance to those of a hypothetical streaming tool based on PaToH. That is the hypergraphs are partitioned by PaToH (with SPEED and DEFAULT configurations) and the run times, cuts (connectivities), and memory usages are normalized with respect to those of {\sc MinMax-n2P} for graphs {\tt coPapersDBLP}, {\tt holywood2009}, {\tt socLiveJournal1}, {\tt com-Orkut}, and {\tt uk2005}.  The experiments show that on these graphs, the DEFAULT configuration can be $10$--$30\%$ better in terms of partitioning quality. However, it can also be more than $100\times$ slower (see the run time ratio bar for {\tt SocLiveJournal1}). Furthermore, PaToH uses $12$--$18\times$ more memory compared to {\sc MinMax-n2P}. Note that PaToH, or any other streaming partitioner, is not suitable for the streaming setting. The results are only given that there is room for improvement on {\tt MinMax-n2p} especially in terms of partitioning quality which rationalizes the attempts for refining the partitioning throughout streaming. 

\begin{table}[htbp]
\resizebox{.45\textwidth}{!}{%
\begin{tabular}{@{}cll|c
>{\columncolor[HTML]{FFFFFF}}c 
>{\columncolor[HTML]{FFFFFF}}c c
>{\columncolor[HTML]{FFFFFF}}c 
>{\columncolor[HTML]{FFFFFF}}c @{}}
\cmidrule(l){4-9}
\multicolumn{1}{l}{} &  &  & \multicolumn{3}{c|}{\cellcolor[HTML]{C0C0C0}Run Time} & \multicolumn{3}{c|}{\cellcolor[HTML]{C0C0C0}Cut} \\ \midrule
\multicolumn{1}{|c}{\cellcolor[HTML]{C0C0C0}Matrix} & \multicolumn{1}{c}{\cellcolor[HTML]{C0C0C0}Buf.} & \multicolumn{1}{c|}{\cellcolor[HTML]{C0C0C0}Passes} & \multicolumn{1}{c|}{\cellcolor[HTML]{C0C0C0}R} & 
\multicolumn{1}{c|}{\cellcolor[HTML]{C0C0C0}RR} & 
\multicolumn{1}{c|}{\cellcolor[HTML]{C0C0C0}RRS} & 
\multicolumn{1}{c|}{\cellcolor[HTML]{C0C0C0}R} & 
\multicolumn{1}{c|}{\cellcolor[HTML]{C0C0C0}RR} & 
\multicolumn{1}{c|}{\cellcolor[HTML]{C0C0C0}RRS} 
\\ \midrule
%\multicolumn{1}{c|}{\cellcolor[HTML]{FFFFFF}} &  & 2 & 2.8 & 4.4 & \multicolumn{1}{c|}{\cellcolor[HTML]{FFFFFF}2.0} & 0.86 & 0.86 & 0.85 \\
%\multicolumn{1}{c|}{\cellcolor[HTML]{FFFFFF}} &  & 4 & 3.9 & 6.4 & \multicolumn{1}{c|}{\cellcolor[HTML]{FFFFFF}2.6} & 0.85 & 0.84 & 0.84 \\
%\multicolumn{1}{c|}{\cellcolor[HTML]{FFFFFF}} & \multirow{-3}{*}{0.05} & 8 & 6.0 & 10.2 & \multicolumn{1}{c|}{\cellcolor[HTML]{FFFFFF}3.8} & 0.85 & 0.82 & 0.83 \\ \hhline{|~|--|}
\multicolumn{1}{c|}{\cellcolor[HTML]{FFFFFF}} &  & 2 & 2.8 & 4.5 & \multicolumn{1}{c|}{\cellcolor[HTML]{FFFFFF}2.1} & 0.84 & 0.82 & 0.87 \\
\multicolumn{1}{c|}{\cellcolor[HTML]{FFFFFF}} &  & 4 & 3.8 & 6.6 & \multicolumn{1}{c|}{\cellcolor[HTML]{FFFFFF}2.8} & 0.83 & 0.81 & 0.87 \\
\multicolumn{1}{c|}{\multirow{-3}{*}{\cellcolor[HTML]{FFFFFF}coPapers}} & \multirow{-3}{*}{0.15} & 8 & 6.0 & 10.8 & \multicolumn{1}{c|}{\cellcolor[HTML]{FFFFFF}4.2} & 0.83 & 0.79 & 0.86 \\ \midrule
%\multicolumn{1}{c|}{\cellcolor[HTML]{FFFFFF}} &  & 2 & 4.0 & 4.4 & \multicolumn{1}{c|}{\cellcolor[HTML]{FFFFFF}3.8} & 0.97 & 0.98 & 0.98 \\
%\multicolumn{1}{c|}{\cellcolor[HTML]{FFFFFF}} &  & 4 & 5.9 & 6.4 & \multicolumn{1}{c|}{\cellcolor[HTML]{FFFFFF}5.5} & 0.97 & 0.98 & 0.98 \\
%\multicolumn{1}{c|}{\cellcolor[HTML]{FFFFFF}} & \multirow{-3}{*}{0.05} & 8 & 9.7 & 10.3 & \multicolumn{1}{c|}{\cellcolor[HTML]{FFFFFF}9.8} & 0.97 & 0.98 & 0.98 \\ \hhline{|~|--|}
\multicolumn{1}{c|}{\cellcolor[HTML]{FFFFFF}} &  & 2 & 4.0 & 4.4 & \multicolumn{1}{c|}{\cellcolor[HTML]{FFFFFF}3.8} & 0.97 & 0.96 & 0.97 \\
\multicolumn{1}{c|}{\cellcolor[HTML]{FFFFFF}} &  & 4 & 5.8 & 6.4 & \multicolumn{1}{c|}{\cellcolor[HTML]{FFFFFF}5.6} & 0.97 & 0.95 & 0.96 \\
\multicolumn{1}{c|}{\multirow{-3}{*}{\cellcolor[HTML]{FFFFFF}hollywood}} & \multirow{-3}{*}{0.15} & 8 & 9.6 & 10.2 & \multicolumn{1}{c|}{\cellcolor[HTML]{FFFFFF}8.8} & 0.96 & 0.95 & 0.97 \\ \midrule
%\multicolumn{1}{c|}{\cellcolor[HTML]{FFFFFF}} &  & 2 & 3.8 & \cellcolor[HTML]{FFFFFF}3.8 & \multicolumn{1}{c|}{\cellcolor[HTML]{FFFFFF}3.4} & 0.96 & 0.95 & 0.95 \\
%\multicolumn{1}{c|}{\cellcolor[HTML]{FFFFFF}} &  & 4 & 5.6 & \cellcolor[HTML]{FFFFFF}5.5 & \multicolumn{1}{c|}{\cellcolor[HTML]{FFFFFF}4.9} & 0.96 & 0.95 & 0.95 \\
%\multicolumn{1}{c|}{\cellcolor[HTML]{FFFFFF}} & \multirow{-3}{*}{0.05} & 8 & 9.1 & \cellcolor[HTML]{FFFFFF}9.0 & \multicolumn{1}{c|}{\cellcolor[HTML]{FFFFFF}7.8} & 0.96 & 0.95 & 0.95 \\ \hhline{|~|--|}
\multicolumn{1}{c|}{\cellcolor[HTML]{FFFFFF}} &  & 2 & 3.8 & 3.8 & \multicolumn{1}{c|}{\cellcolor[HTML]{FFFFFF}3.4} & 0.96 & 0.94 & 0.95 \\
\multicolumn{1}{c|}{\cellcolor[HTML]{FFFFFF}} &  & 4 & 5.5 & 5.7 & \multicolumn{1}{c|}{\cellcolor[HTML]{FFFFFF}4.9} & 0.96 & 0.94 & 0.94 \\
\multicolumn{1}{c|}{\multirow{-3}{*}{\cellcolor[HTML]{FFFFFF}soc-Live}} & \multirow{-3}{*}{0.15} & 8 & 9.0 & 9.3 & \multicolumn{1}{c|}{\cellcolor[HTML]{FFFFFF}7.8} & 0.96 & 0.94 & 0.94 \\ 
\end{tabular}%
}
\caption{\small{Effect of the number of $passes$ on refinement algorithms; the results are averaged for $K = 256$ and $K = 2048$. The experiments are executed on the \textbf{Server}. All of the values are normalized with respect to those of {\sc MinMax} on the same experiment.}}
\label{tab:passes}
\end{table}

To determine the optimal number of $passes$ for the refinement-based algorithms, we performed 2, 4, and 8 passes over the buffered vertices and measure the runtime and partitioning quality of \mmref, \mmrefrlx, and \mmrefrlxsv. Table~\ref{tab:passes} presents the results of these experiments with the a buffer capacity $B = 0.15 \times |\Hy|$ for each hypergraph $|\Hy|$. The results show that although refinement can be useful for reducing the connectivity, its overhead is significant. Furthermore, after $2$ passes, there is only a minor improvement on the partitioning quality. Still, using $passes = 4$ can reduce the cut size for a lot of experiments which is a decision we follow in the rest of the experiments.

\begin{figure*}
     \begin{subfigure}[htbp]{0.5\textwidth}
         \centering
         \includegraphics[width=\textwidth]{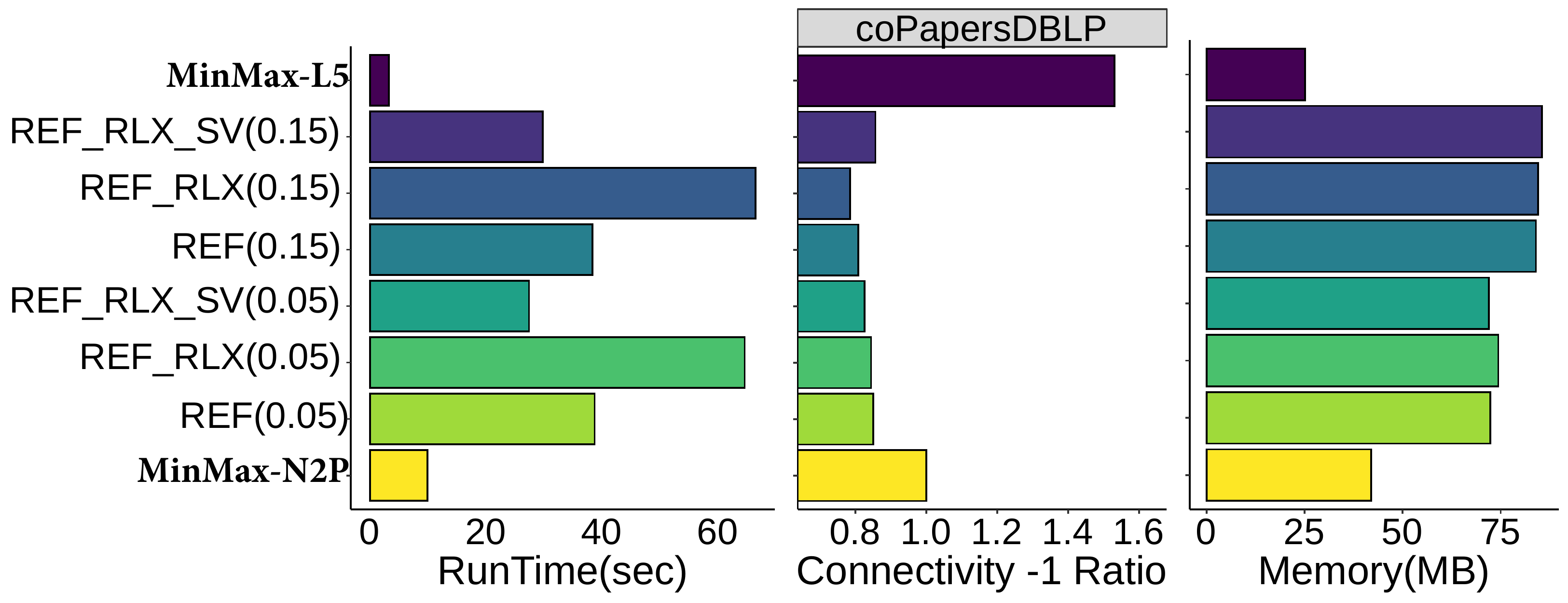}
         \label{fig:cop}
     \end{subfigure}
     \begin{subfigure}[htbp]{0.5\textwidth}
         \centering
         \includegraphics[width=\textwidth]{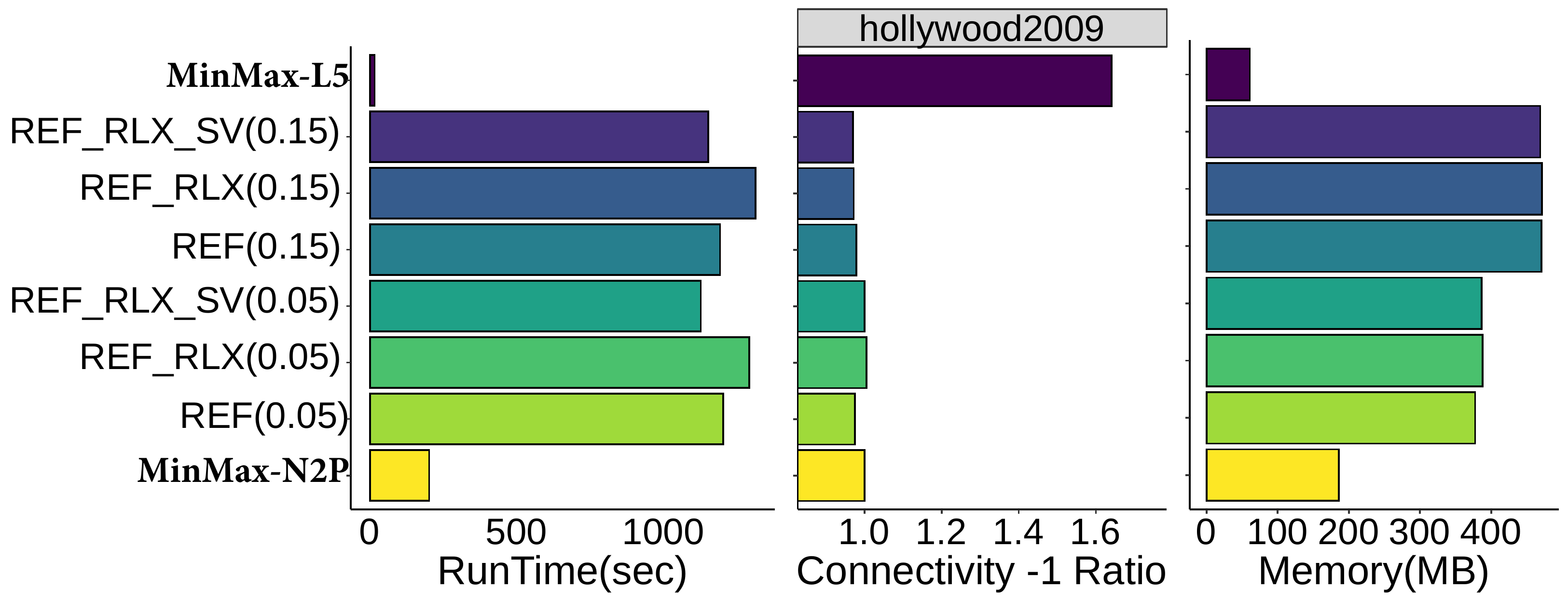}
         \label{fig:hol}
     \end{subfigure}
     
     \begin{subfigure}[htbp]{0.5\textwidth}
         \centering
         \includegraphics[width=\textwidth]{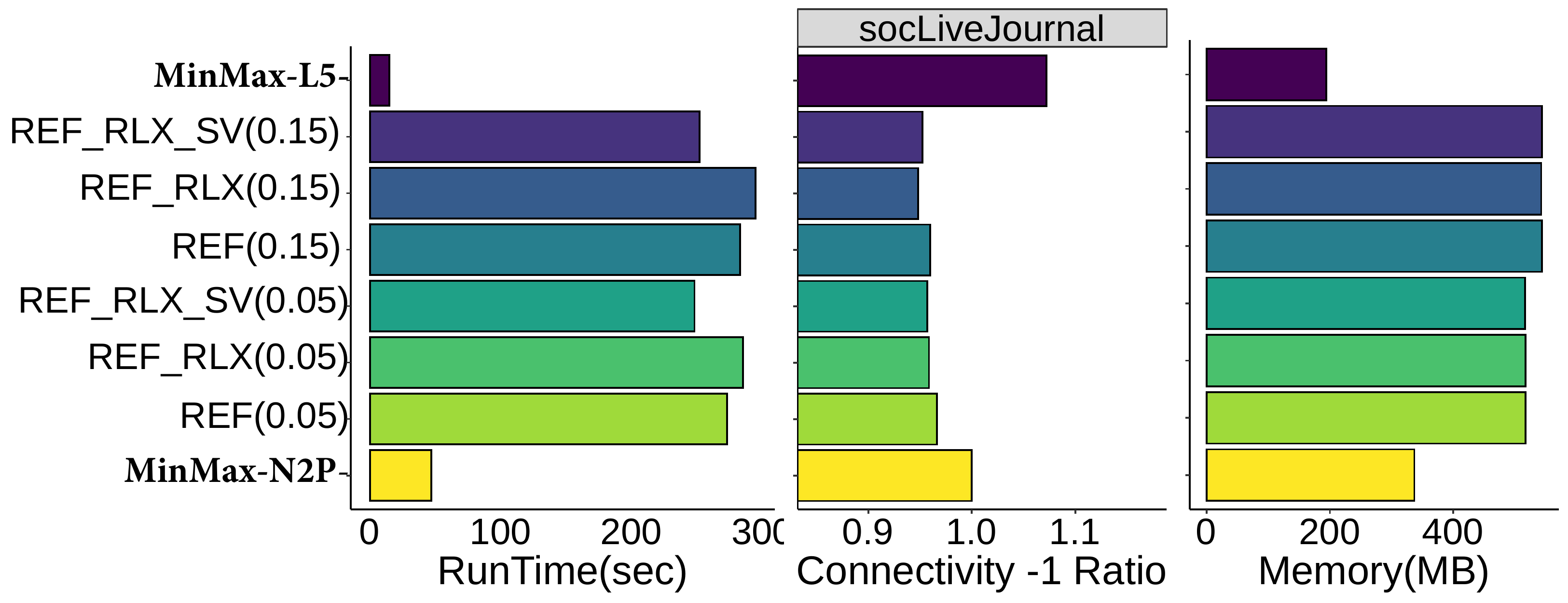}
         \label{fig:soc}
     \end{subfigure}
     \hfill
       \begin{subfigure}[htbp]{0.5\textwidth}
         \centering
         \includegraphics[width=\textwidth]{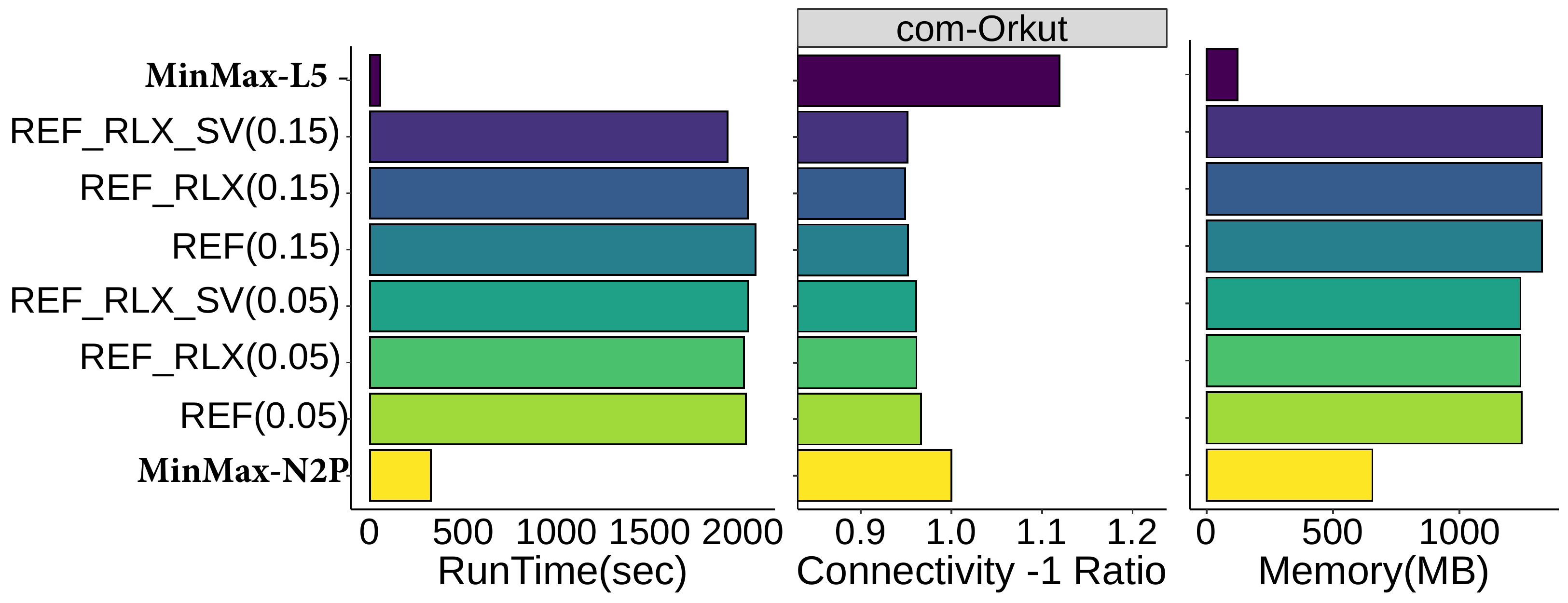}
         \label{fig:ork}
     \end{subfigure}

     \begin{subfigure}[htbp]{0.5\textwidth}
         \centering
         \includegraphics[width=\textwidth]{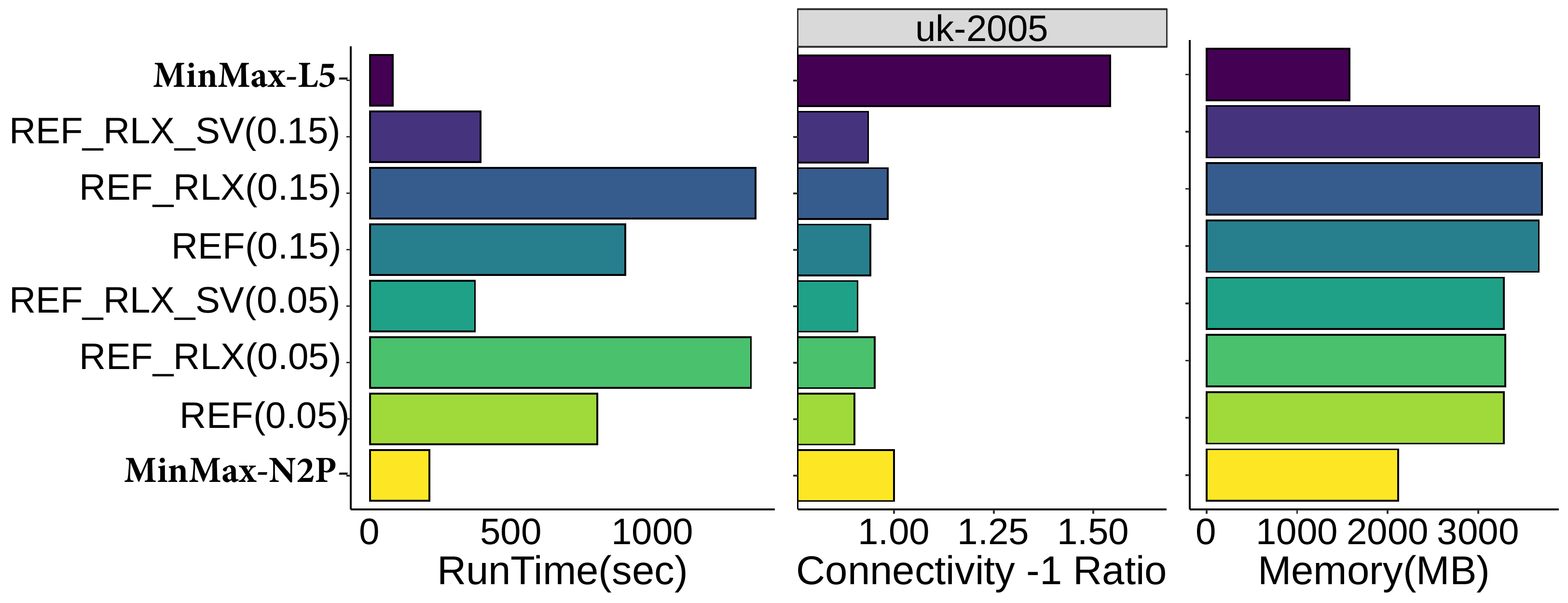}
         \label{fig:ukw}
     \end{subfigure}
       %\begin{subfigure}[htbp]{0.5\textwidth}
        % \centering
        % \includegraphics[width=\textwidth]{figures/sk_r49}
        % \label{fig:skw}
     %\end{subfigure}
        \begin{subfigure}[htbp]{0.5\textwidth}
        \centering
         \includegraphics[width=\textwidth]{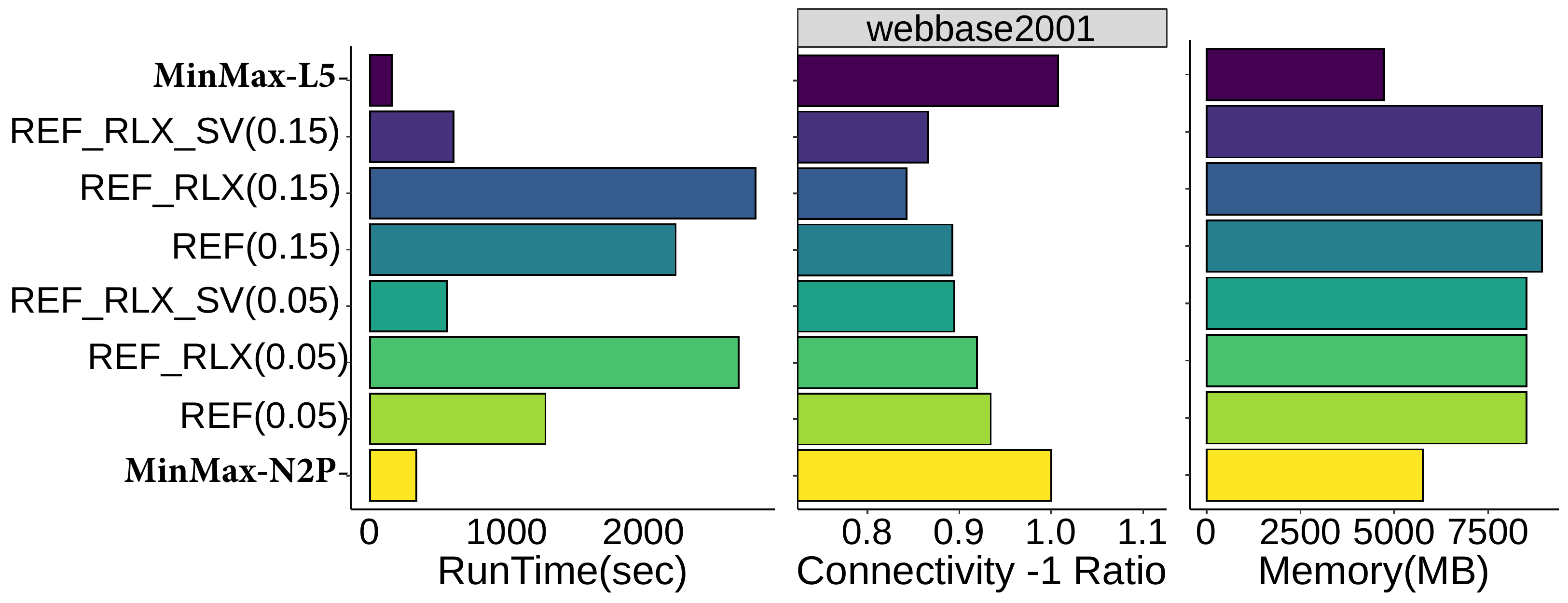}
         \label{fig:web}
     \end{subfigure}
     
        \caption{\small{The run times (in seconds), connectivity scores (normalized with respect to {\sc MinMax-n2p}), and memory usages (in MBs) of the refinement heuristics, {\sc MinMax-n2p}, and {\sc MinMax-L5} on all hypergraphs. The experiments are executed on the {\bf Server} for $K = 2048$. An imbalance ratio of $0.1$  is used for all experiments and $passes = 4$ is used for the refinement heuristics. In the figure, the parameter for these heuristics shows the parameter $\theta$ to find the buffer size $B = \theta \times |\Hy|$ for each hypergraph $\Hy$.}}
        \label{fig:mainexp}
\end{figure*}

Figure~\ref{fig:mainexp} shows the run times (in seconds), connectivity scores (normalized with respect to {\sc MinMax-n2p}), and memory usages (in MBs) of the refinement heuristics, {\sc MinMax-n2p}, and {\sc MinMax-L5} on all hypergraphs in Table~\ref{tab:graph_shape}. The experiments are executed on the {\bf Server}. An imbalance ratio of $0.1$ and $passes = 4$ are used for the experiments. As the results show, the refinement based heuristics improve the partitioning quality in between $5$--$20\%$ depending on the hypergraph. Furthermore, when the buffer size is increased, these heuristics tend to improve the quality better. Besides, for the two largest graphs in our experiments, REF\_RLX\_SV is much faster than the other two refinement heuristics REF and REF\_RLX with a similar improvement over the partitioning quality and the same memory usage. Hence, it can be a good replacement to the original {\sc MinMax-n2p} with no refinement if the partitioning quality has the utmost importance. 

%\lipsum[9-12]

%\begin{figure}[htbp]
%    \centering
%    \includegraphics[width=.35\textwidth]{figures/cartesian_3.pdf}
%    \caption{The performance of the refinement algorithms in terms of run time and memory as axes and cut as diameter. Values are normalized with respect to \mmnew and diameters are visualized as $cut^7$ in favor of readability.}
%    \label{fig:cartesian}
%\end{figure}

\begin{table}[htbp]
\resizebox{.45\textwidth}{!}{%
\begin{tabular}{@{}llrrrr@{}}
 & \multicolumn{1}{l|}{} & \multicolumn{4}{c|}{\cellcolor[HTML]{9B9B9B}Parts} \\ \midrule
\rowcolor[HTML]{EFEFEF} 
\multicolumn{1}{l}{\cellcolor[HTML]{9B9B9B}Device} & Algorithm & 32 & 256 & 2048 & 16384 \\ \midrule
\multicolumn{1}{l|}{\cellcolor[HTML]{C0C0C0}} & {\sc MinMax-n2p} & 142.6 & 1.43$\times$ & 2.16$\times$ & 3.73$\times$ \\
\multicolumn{1}{l|}{\cellcolor[HTML]{C0C0C0}} & REF & 593.1 & 1.64$\times$ & 2.93$\times$ & 4.38$\times$ \\
\multicolumn{1}{l|}{\cellcolor[HTML]{C0C0C0}} & REF\_RLX & 650.0 & 1.50$\times$ & 2.89$\times$ & 4.16$\times$ \\
\multicolumn{1}{l|}{\cellcolor[HTML]{C0C0C0}} & REF\_RLX\_SV & 532.3 & 1.53$\times$ & 2.89$\times$ & 4.73$\times$ \\
\multicolumn{1}{l|}{\multirow{-6}{*}{\cellcolor[HTML]{C0C0C0}Pi-1GB}} & {\sc MinMax-L5} & 55.6 & 1.17$\times$ & 1.30$\times$ & 1.74$\times$ \\ \midrule
\multicolumn{1}{l|}{\cellcolor[HTML]{C0C0C0}} & {\sc MinMax-n2p} & 72.4 & 1.32$\times$ & 1.81$\times$ & 2.80$\times$ \\
\multicolumn{1}{l|}{\cellcolor[HTML]{C0C0C0}} & REF & 335.2 & 1.41$\times$ & 2.32$\times$ & 3.62$\times$ \\
\multicolumn{1}{l|}{\cellcolor[HTML]{C0C0C0}} & REF\_RLX & 362.8 & 1.33$\times$ & 2.27$\times$ & 3.60$\times$ \\
\multicolumn{1}{l|}{\cellcolor[HTML]{C0C0C0}} & REF\_RLX\_SV & 287.5 & 1.47$\times$ & 2.37$\times$ & 3.92$\times$ \\
\multicolumn{1}{l|}{\multirow{-6}{*}{\cellcolor[HTML]{C0C0C0}Odroid-2GB}} & {\sc MinMax-L5} & 36.1 & 1.11$\times$ & 1.19$\times$ & 1.40$\times$ \\ \midrule
\multicolumn{1}{l|}{\cellcolor[HTML]{C0C0C0}} & {\sc MinMax-n2p} & 71.5 & 1.34$\times$ & 1.89$\times$ & 2.85$\times$ \\
\multicolumn{1}{l|}{\cellcolor[HTML]{C0C0C0}} & REF & 308.3 & 1.45$\times$ & 2.46$\times$ & 3.73$\times$ \\
\multicolumn{1}{l|}{\cellcolor[HTML]{C0C0C0}} & REF\_RLX & 339.8 & 1.35$\times$ & 2.43$\times$ & 3.62$\times$ \\
\multicolumn{1}{l|}{\cellcolor[HTML]{C0C0C0}} & REF\_RLX\_SV & 264.8 & 1.52$\times$ & 2.57$\times$ & 4.05$\times$ \\
\multicolumn{1}{l|}{\multirow{-6}{*}{\cellcolor[HTML]{C0C0C0}LattePanda-4GB}} & {\sc MinMax-L5} & 30.4 & 1.11$\times$ & 1.18$\times$ & 1.33$\times$ \\ 

%\multicolumn{1}{l|}{\cellcolor[HTML]{C0C0C0}} & {\sc MinMax-n2p} & 63.1 & 1.34$\times$ & 1.86$\times$ & 3.00$\times$ \\
%\multicolumn{1}{l|}{\cellcolor[HTML]{C0C0C0}} & REF & 273.7 & 1.49$\times$ & 2.42$\times$ & 3.81$\times$ \\
%\multicolumn{1}{l|}{\cellcolor[HTML]{C0C0C0}} & REF\_RLX & 297.1 & 1.38$\times$ & 2.39$\times$ & 3.76$\times$ \\
%\multicolumn{1}{l|}{\cellcolor[HTML]{C0C0C0}} & REF\_RLX\_SV & 243.0 & 1.51$\times$ & 2.49$\times$ & 4.13$\times$ \\
%\multicolumn{1}{l|}{\multirow{-6}{*}{\cellcolor[HTML]{C0C0C0}Jetson-4GB}} & {\sc MinMax-L5} & 27.0 & 1.13$\times$ & 1.23$\times$ & 1.46$\times$ \\
\end{tabular}%
}
\caption{\small{Run times of the proposed algorithms on single board devices and {\tt socLiveJournal1} for different $K$ values. The allowed imbalance is set to $0.1$, and the buffer capacity $B = 0.15 \times |\Hy|$ is used. The results on the $K = 32$ column are given in seconds, and for $K = 256, 2048$ and $16384$, the results are normalized with respect to $K = 32$. That is the results in the last three columns show the decrease in the run time performance when $K$ is increased from $32$ to the corresponding value.}}
\label{tab:vsnk1}
\end{table}

\balance
Table~\ref{tab:vsnk1} shows the run time performance of the proposed algorithms on different architectures and for $K = 32, 256, 2048$ and $16384$. The hypergraph {\tt socLiveJournal1} is used for these experiments. The results are similar to the ones in the {\bf Server}. Yet additionally, the slow-down values in the last three columns show that using much less memory, {\sc MinMax-L5} stays more scalable compared to other algorithms when $K$ is increased. Furthermore, the overhead of the refinement heuristics deteriorates the scaling behavior when they are added on top of the {\sc MinMax-n2p}. However, their negative impact tends to decrease when an SBC with more memory is used. This also shows the importance of streaming hypergraph algorithms with low-memory footprints in practice. 

\section{Conclusion and Future Work} \label{sec:con}

In this work, we focused on the streaming hypergraph partitioning problem. The problem has unique challenges compared to similar problems in the streaming setting such as graph partitioning. We significantly improved the run time performance of a well-known streaming algorithm {\sc MinMax} and proposed several variants on top of it to reduce the memory footprint and improve the partitioning quality. The experiments show that there is still room for improvement for these algorithms. As future work, we are planning to devise more advanced techniques that can overcome the trade-off among the run time, memory usage, and partitioning quality.

\section*{Acknowledgements} This work is funded by The Scientific and Technological Research Council of Turkey (T\"{U}BITAK) under the grant number 117E249.

\bibliographystyle{IEEEtran}
\bibliography{references}

\newpage

%\includepdf[page={1}]{crystk03.pdf}
%\includepdf[page={1}]{wikitalk.pdf}
%\includepdf[page={1}]{dgreen.pdf}
%\includepdf[page={1}]{amazon.pdf}KaHyPar
%\includepdf[page={1}]{nemeth.pdf}

\end{document}